\documentclass[preprint,pre,aps,showkeys,nofootinbib,amsmath,amssymb,floatfix]{revtex4-2}
\usepackage{enumerate}
\usepackage{amsmath}
\usepackage{url}
\usepackage{amssymb}
\usepackage{graphicx}
\usepackage{dcolumn}
\usepackage{bm}
\usepackage{float}
\usepackage{placeins}
\usepackage{subcaption}
\usepackage[usenames,dvipsnames]{color}

\begin{document}

\preprint{APS/123-QED}

\title{Conditional Residence Times and Sequential Transition Dynamics of an Overdamped Dimer}

\author{Dhruv Agrawal}
 \altaffiliation[Also at ]{Department of Physics, National Institute of Technology Meghalaya}
\author{W. L. Reenbohn }%
 \email{wlreenbohn@nitm.ac.in}
\affiliation{%
 Department of Physics, National Institute of Technology Meghalaya, Saitsohpen Sohra, Meghalaya 793108, India
}%



\date{\today}

\begin{abstract}
 We investigate the completion dynamics of an overdamped dimer moving in a bistable potential under thermal fluctuations and a weak periodic force. Both monomers start in one of the two wells separated by a barrier. The transition is initiated when the monomer closer to the barrier makes a jump across it. The completion dynamics refers to the next part of the dynamics where the second monomer has to wait for some time before it can follow up. We use the Conditional Residence Time (CRT) to study the delay between the successive barrier crossing of the two monomers. The CRT distributions highlight qualitatively different regimes formed by the competition between the escape times of the lagging monomer and the time period of the external drive. The effect is strongest in the weak coupling regime where the delayed completion is spread across multiple forcing cycles. By partitioning this process into three windows, i.e. the immediate, first cycle and later cycles, we show that the probability that the lagging monomer will make a transition in the said cycle is redistributed among these pathways as we change the frequency of the drive. This leads to a non-monotonic dependence of the mean CRT on the frequency of the drive. Our results demonstrate that transition initiation and completion in a coupled system are two separate processes and establish CRT as a useful measure to quantify the sequential barrier crossing dynamics in coupled stochastic systems.
\end{abstract}


\maketitle

\section{\label{sec1}Introduction \protect}
Noise-driven transitions across energy barriers occur commonly across various fields of physics, chemistry and biology, i.e., from molecular conformational changes and enzyme kinetics to signal detection by neurons and paleo-climatic variability \cite{10.1098/rsos.171531, PhysRevLett.122.158701,tripathi2022acceleration,PhysRevE.89.032139, guo2018functional}. Thermally activated escape dynamics was classically described by Kramers in his reaction rate theory, linking transition rates across the barrier to the corresponding barrier heights and noise strength. The theory provides the basic exponential picture of residence times in metastable wells \cite{KRAMERS1940284, RRT}.

Noise-induced transitions are often characterized by quantities like the escape rate, mean first passage times (MFPTs), survival probabilities and residence time distributions (RTDs). The quantifiers provide information about the likelihood and the timing associated with the noise-induced transitions \cite{quack1981van, gardiner2009handbook, elber2019molecular}. 
The work on noise induced transitions was extended further by adding a weak periodic forcing. The studies in \cite{cc1981,cc1982,cc1983} established that an optimum tuning of the noise strength with the frequency of the periodic input enhances the response of the system. This constructive role of noise was termed as stochastic resonance (SR). Since then SR has been extensively studied in different kinds of settings and has been verified experimentally in many contexts. It provides a unifying framework for understanding how noise, non-linearity and periodic modulation interact together to produce a coherent switching dynamics \cite{Mcnamara1989,douglass1993noise,Gamm1998}. Initial studies on SR were focused on single degrees of freedom, where the residence time distribution (RTD) has been an important quantifier. The residence time distribution (RTD) refers to the statistics of the amount of time that the system spends within a metastable state before making a noise-assisted transition to the alternate state. This provides a direct insight into the barrier-crossing kinetics under a non-equilibrium driving. In the classical overdamped single-particle system, Kramers-type escape yields an exponential tail in the RTD, indicating a memory-less Poisson process. This process occurs at an effective rate which depends on the barrier height and thermal noise level. This scenario is repeatedly validated in bistable systems exhibiting random transitions between metastable states. Such Poissonian residence statistics have been explicitly demonstrated in extended interacting systems where macroscopic transitions reduce to rare-event activated kinetics with Arrhenius-like scaling of lifetimes, reinforcing the exponential long-time behavior of the RTD in the Markovian regime \cite{PhysRevE.89.032139}. A time-periodic external drive generically imprints further structure onto the RTD by periodically modulating the effective barrier, enhancing escape probability in specific phases of the cycle and suppressing it in others. Near stochastic-resonance conditions, this phase sensitivity leads to concentration of residence events at half-integer multiples of the forcing period, producing distinct peaks in the RTD that signal phase locking between noise-activated escape and the drive. Such peaked residence-time patterns have been reported as a hallmark of stochastic resonance in bistable and excitable systems. Theoretical analysis of RTDs in driven bistable systems have clarified how time dependent rates influence the structure of the RTD, modifying it between an exponentially tailed distribution and a multi-peak structure \cite{rtd2003,rtd2004,Talkner_2005}.
Although, these rate based measures are widely used, recent studies have shown a growing interest in measures related to the times associated with specific regions of the trajectory which actually lead to transitions across the barrier \cite{elber2019molecular,10.1063/1.4936408, doi:10.1021/acs.jpcb.9b01616, doi:10.1073/pnas.2008307117}. The statistics related to such transition path times provides valuable insights into the dynamical characteristics of the multistable system.

Most of the earlier studies use a single particle in their model. The quantifiers which exist are developed keeping this single particle picture in mind. For a single particle that is trapped in some potential well, the escape from the well across a finite barrier involves the particle spending a significant amount of time at the bottom of the well and a relatively fast transition across the barrier.
However, real systems are seldom single isolated particles but a collection of several interacting subunits. Interactions with similar particles produce collective effects that influence both the initiation and the completion of transitions \cite{kenfack2010,asfaw2012,PhysRevE.96.032108,LYNGDOH2024,62t9-g2n6}. Dimers and small chain polymers serve as basic models for studying cooperative escape dynamics in polymer physics, molecular motors, and experimental platforms where few degrees of freedom are coupled and driven by noise and external fields. Coupling is known to either suppress or enhance the coordinated switching when tuned appropriately. The escape process involves multiple intermediate stages and hence cannot be effectively studied by single particle description. The simplest model that can be considered is that of a dimer, i.e., two subunits coupled by a spring \cite{kenfack2010, asfaw2012}, short range interactions or a combination of both \cite{LYNGDOH2024, 62t9-g2n6}. In such a system, the transition is said to be initiated when the monomer closer to the barrier, i.e. the leading monomer, crosses the barrier first, while the second monomer, i.e. the follower, often waits before making the transition. The transition of the complete dimer may fail if the leader returns to the original well before the follower completes the transition. The motion is often simplified by considering the center of mass coordinate of the system under study. In \cite{62t9-g2n6}, we use the center of mass coordinate of the dimer to show the SR phenomena using different quantifiers like hysteresis loop area, input energy per period etc. We highlight there another important aspect of the transition process concerning coupled systems, i.e., the successful transition ratio, which again remains hidden if we use the center of mass coordinate. This serves as a good example to show that although the center of mass serves as a simplified coordinate to study coupled systems, it compromises with the intricacies of the motion of the individual monomers.

In this work, we address this gap further by introducing the Conditional Residence Time (CRT), defined as the time delay between the successive crossing of the leader and the follower. Unlike the conventional residence times which measures the waiting period preceding a transition, CRT focuses on the dynamics occurring immediately after the transition is initiated. CRT can be interpreted as a conditional first passage time associated with coupled systems. It focuses on the temporal coordination between interacting parts of a coupled system and measures how rapidly an initiated transition leads to a successful one.
 

The importance of treating transition initiation and completion distinctly is related to concepts emerging from transition path theory, where the highlight shifts from focusing on the probability of escape to the dynamics occurring solely during the transition. However, most of the existing work on escape of coupled particles uses global quantifiers like the mean residence times. A systematic study on the delay associated with the transition completion in coupled systems remains largely unexplored.

In this work, we present an investigation of the conditional barrier-crossing dynamics of an overdamped dimer in a bistable potential. The monomers interact via the sum of harmonic and Lennard-Jones potential. The system is subjected to an external periodic force and thermal noise. We quantify the conditional dynamics using CRT. The CRT distributions reveal dynamical regimes which are ignored by conventional measures. CRTDs show that the competition between the escape timescales and the driving period generates different completion pathways. These include nearly memoryless follower dynamics and transitions that extend over several cycles of the drive. By strictly focusing on the completion part of the dynamics, CRT offers a framework for characterizing the sequential transition dynamics in coupled systems.
\section{\label{sec2}The Model \protect}
\noindent We consider a system of an overdamped dimer composed of two subunits or monomers moving in a one-dimensional double well potential $V(x)$ given by, 
\begin{equation}\label{BP}
    V(x) = -\frac{\omega_B^2}{2}x^2 + \frac{\omega_B^2}{4x_m^2}x^4
\end{equation}
where, the potential minima are located at $\pm x_m$, the barrier height is $V_B=x_m^2\omega_B^2/4$ and $\omega_B^2=\omega_0^2/2$. $\omega_B^2$ and $\omega_0^2$ are the curvature of the potential at the maxima and minima of the potential respectively. The length of the dimer is $r= \lvert x_1-x_2 \rvert$. Then, the interaction potential and the forces between the monomers is given by the combination of:
\begin{enumerate}
    \item Harmonic potential: 
    \begin{equation}\label{hp}
        V_H(r) = \frac{1}{2}k(r - l_0)^2
    \end{equation}
    \begin{equation}
    F_H = -k(r-l_0)
\end{equation}
    where $k$ and $l_0$ are the coupling strength and the equilibrium length of the dimer. 
    \item Lennard-Jones (LJ) potential:
    \begin{equation}\label{lp}
    V_{LJ}(r) = 4\epsilon\left(\left( \frac{\sigma}{r}\right)^{12}-\left(\frac{\sigma}{r}\right)^{6} \right)
    \end{equation}
    \begin{equation}
    F_{LJ} = 4\epsilon\left(12\left( \frac{\sigma^{12}}{r^{13}}\right)-6\left(\frac{\sigma^6}{r^7}\right) \right)
\end{equation} 
    where $\epsilon$ is the depth of the LJ potential and $\sigma = 2^{-1/6}l_0$.
\end{enumerate}
The time evolution of this coupled system, when subjected to zero mean Gaussian white noise $(\eta_i(t))$, is described by the following coupled Langevin equations:



\begin{align}
    \gamma\frac{\mathrm{d}x_1}{\mathrm{d}t} = -V'(x_1) + F_H(r)+ F_{LJ}(r) + \eta_1(t) 
\end{align}

\begin{align}
    \gamma\frac{\mathrm{d}x_2}{\mathrm{d}t} = -V'(x_2) - F_H(r)- F_{LJ}(r) + \eta_2(t)
\end{align}
where $x_1$ and $x_2$ are the position coordinates of the left and right particle respectively. In order to investigate SR, an external weak sinusoidal force of magnitude $A_0$ and angular frequency $\Omega$ is added to the system. The equations thereby become,
\begin{equation}\label{eq:fn1}
     \gamma\frac{\mathrm{d}x_1}{\mathrm{d}t} = -V'(x_1) + F_H(r)+ F_{LJ}(r) +A_0\sin{\Omega t}+ \eta_1(t)
\end{equation}

\begin{equation}\label{eq:fn2}
    \gamma\frac{\mathrm{d}x_2}{\mathrm{d}t} = -V'(x_2) - F_H(r)- F_{LJ}(r) +A_0\sin{\Omega t}+ \eta_2(t)
\end{equation}
For numerical efficiency, all variables are converted to dimensionless form by introducing a natural energy scale $V_B$, a length scale $x_m$, and a characteristic time $\tau=\gamma x_m^2/V_B$. Ignoring the bar over each dimensionless quantity, the Langevin equation in dimensionless form becomes,

\begin{equation}\label{dlfn1}
     \frac{\mathrm{d}x_1}{\mathrm{d}t} = 4(x_1 - x_1^3) -4k(r-l_0) + 4\epsilon\left(12\left( \frac{\sigma^{12}}{r^{13}}\right)-6\left(\frac{\sigma^6}{r^7}\right) \right) +A_0\sin{\Omega t}+ \eta_1(t)
\end{equation}

\begin{equation}\label{eq:dlfn2}
    \frac{\mathrm{d}x_2}{\mathrm{d}t} = 4(x_2 - x_2^3) + 4k(r-l_0) - 4\epsilon\left(12\left( \frac{\sigma^{12}}{r^{13}}\right)-6\left(\frac{\sigma^6}{r^7}\right) \right) +A_0\sin{\Omega t}+ \eta_2(t)
\end{equation}
The characteristics of Gaussian white noise are,
\begin{equation}
    \langle \eta_i(t)\rangle = 0
\end{equation}

\begin{equation}
    \langle\eta_i(t)\eta_j(t')\rangle = 2D\delta_{ij}\delta(t-t')
\end{equation}
This model provides a basis for exploring how external periodic drive, thermal fluctuations, and internal coupling, along with excluded volume effects, together describe the barrier crossing dynamics of a dimer.

\section{\label{sec4}Results \protect}
The model has been studied for the following simulation parameters:
\begin{enumerate}
    \item $\Delta t=10^{-3}$
    \item $t=10^6$
    \item $l_0=0.4$
    \item $\epsilon=0.1$
    \item $A_0=0.1$

\end{enumerate}
The above parameters are kept fixed throughout the study. The noise strength ($D$), coupling strength ($k$) and the frequency of the periodic drive ($\Omega$) is varied to analyze the system. Runge-Kutta method of order $2$ is used to solve the coupled differential equations. The results are averaged over a total of $100$ trajectories.

\subsection{Conditional Residence Time}
While the residence time distribution (RTD) characterizes the overall time a particle stays in one well, the description for the case of a coupled system cannot be stated in a similar manner. The dimer undergoes a complete transition across the barrier in two steps. In a coupled system, barrier crossing is typically initiated by one monomer—the leading monomer—which first surmounts the potential barrier. The lagging monomer or the follower influenced by the modified interaction landscape created by the leader’s movement, either follows after a finite delay or the leading monomer is pulled back to the well from where the transition event started. This delay is crucial for understanding the temporal coordination and cooperative nature of transitions in coupled systems.

In order to capture this sequential aspect, we define the Conditional Residence Time (CRT) as the time interval between the instant the leading monomer crosses a prescribed threshold on one side of the barrier and the subsequent crossing of the lagging monomer through the same threshold. The CRT, therefore, quantifies how quickly the system completes a transition once initiated. Unlike the RTD, which measures how long the system resides in a potential well before any transition begins, the CRT isolates the intra-event dynamics—the kinetics of completion following the initiation.

\begin{figure}[H]
    \centering
    
    \begin{subfigure}[b]{0.3\textwidth}
        \centering
        \includegraphics[width=\textwidth]{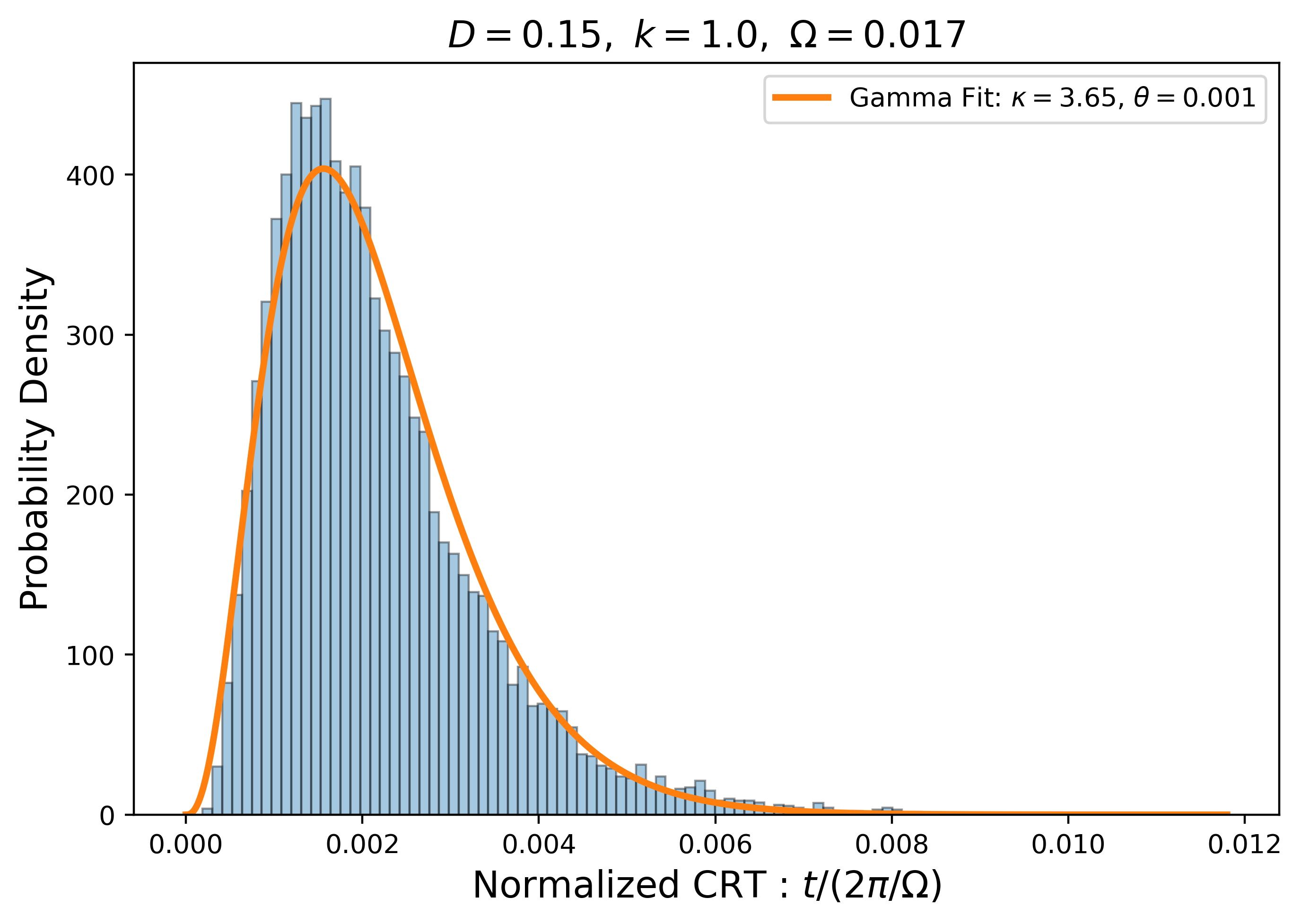}
        \caption{$D=0.15$}
        \label{fig:td2img1}
    \end{subfigure}
    \hfill
    \begin{subfigure}[b]{0.3\textwidth}
        \centering
        \includegraphics[width=\textwidth]{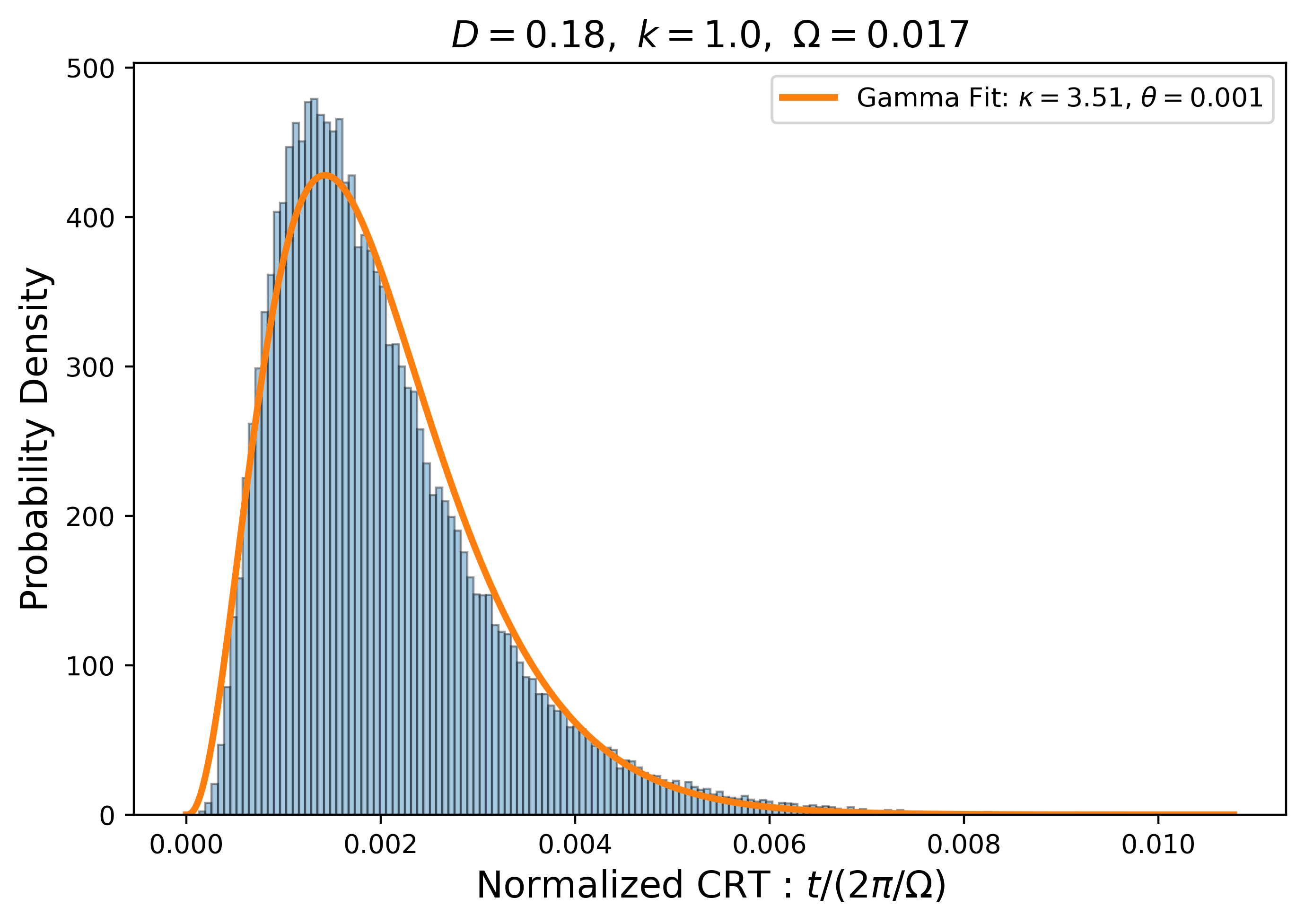}
        \caption{$D=0.18$}
        \label{fig:td2img2}
    \end{subfigure}
    \hfill
    \begin{subfigure}[b]{0.3\textwidth}
        \centering
        \includegraphics[width=\textwidth]{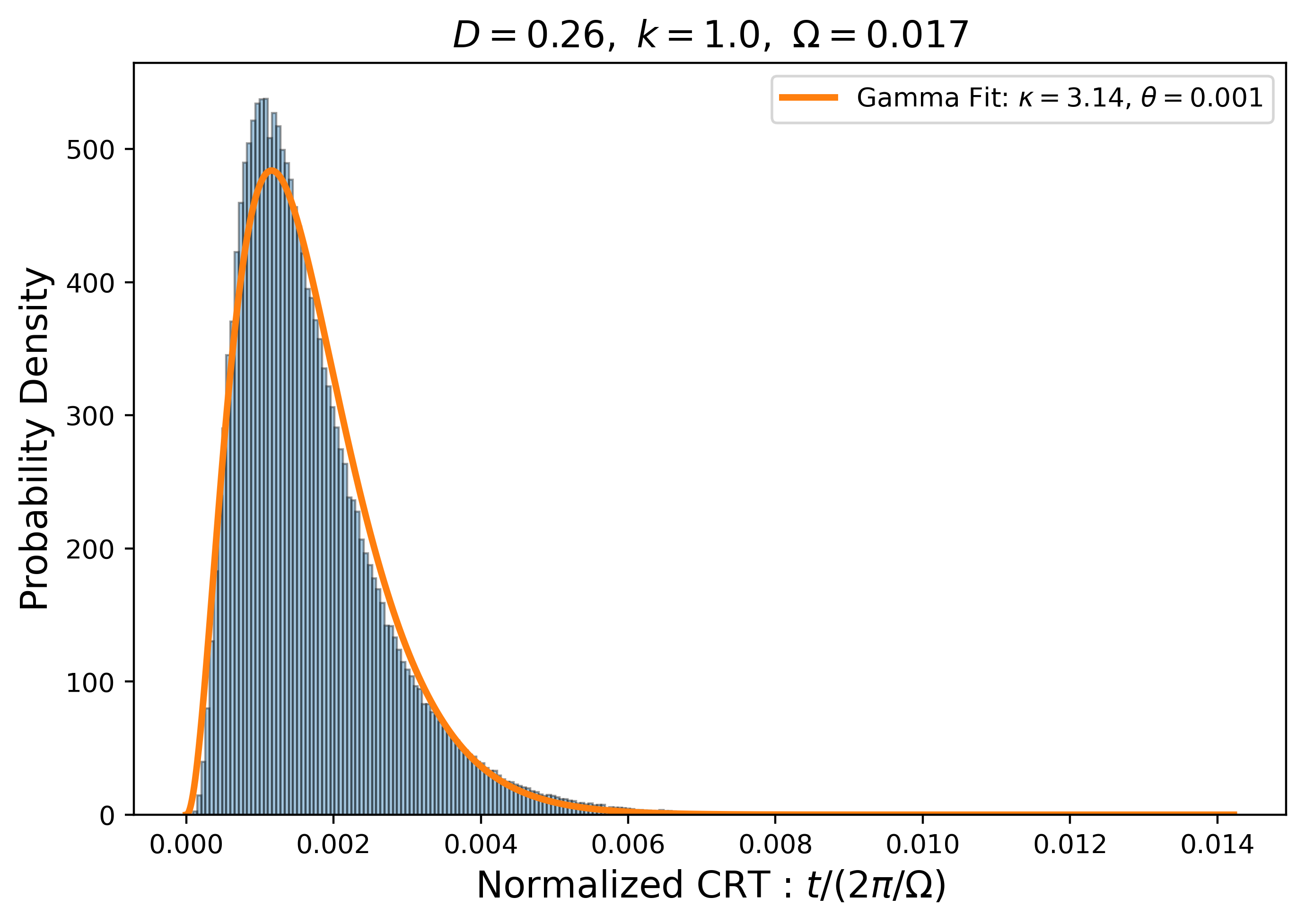}
        \caption{$D=0.26$}
       \label{fig:td2img3}
    \end{subfigure}
    \caption{Conditional residence time distribution for $k=1.0$ and $\Omega=0.017$.}
    \label{fig:t2}
\end{figure}

This measure provides a direct probe of the strength and nature of coupling between the monomers. A narrow CRT distribution indicates highly synchronized motion, as seen in the strong-coupling regime, where the dimer behaves nearly as a single unit. Conversely, a broad CRT distribution signals partial synchronization or intermittent lagging behavior, revealing how the internal interaction, noise, and external forcing jointly determine the catch-up process of the lagging monomer. Fig.\ref{fig:t1}-\ref{fig:t3} show the CRT distributions for different combinations of noise strength, coupling, and frequency of the periodic drive. The distributions are normalized by dividing the measured CRTs by the corresponding time period of the external drive. We start by considering two regimes based on the comparison between the Kramer's escape time $\tau_K$ for the leader and the time period of the drive $T_{\Omega}$:

 \begin{figure}[H]
    \centering
    \begin{subfigure}[b]{0.3\textwidth}
        \centering
        \includegraphics[width=\textwidth]{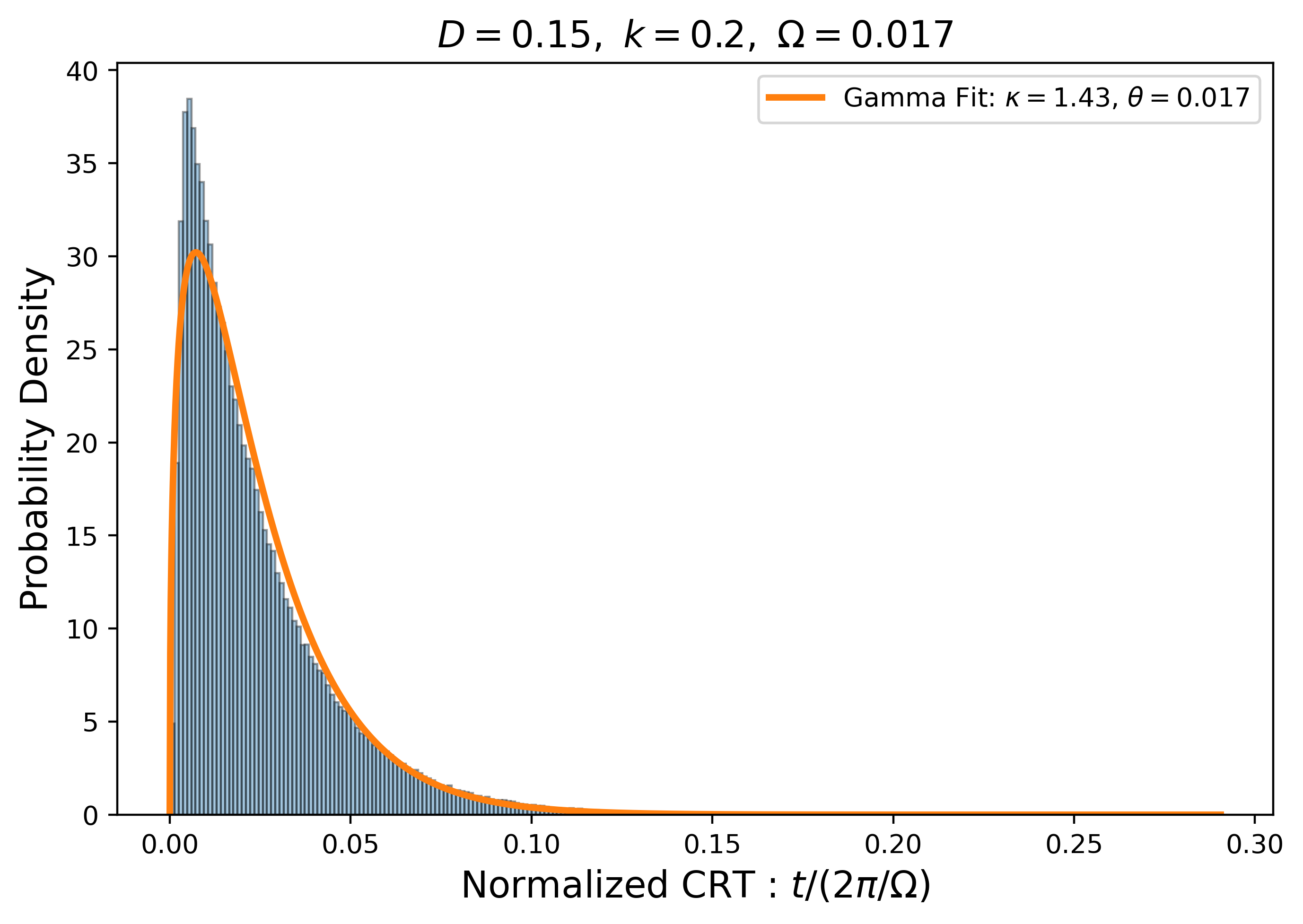}
        \caption{$D=0.15$}
        \label{fig:td1img1}
    \end{subfigure}
    \hfill
    \begin{subfigure}[b]{0.3\textwidth}
        \centering
        \includegraphics[width=\textwidth]{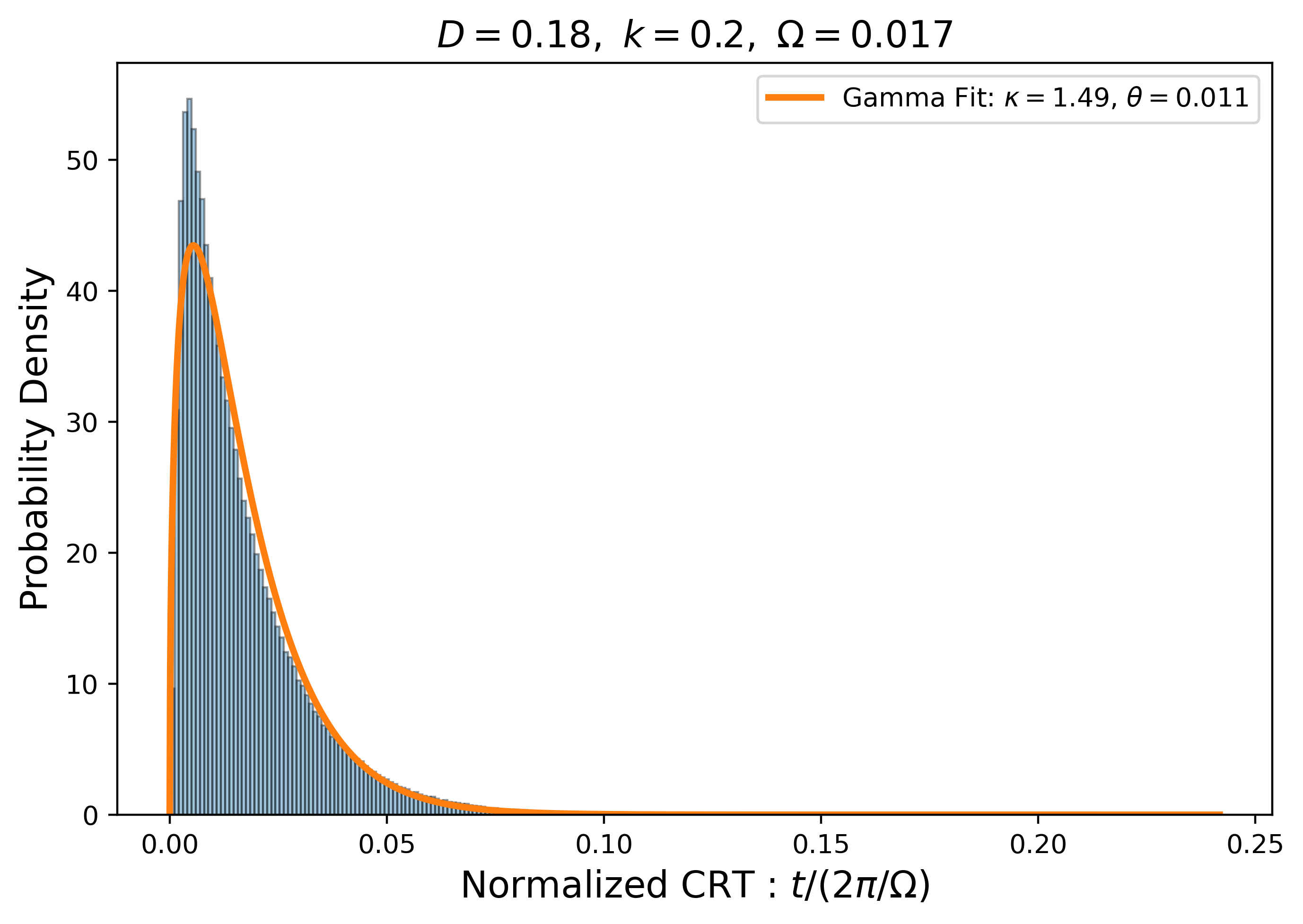}
        \caption{$D=0.18$}
        \label{fig:td1img2}
    \end{subfigure}
    \hfill
    \begin{subfigure}[b]{0.3\textwidth}
        \centering
        \includegraphics[width=\textwidth]{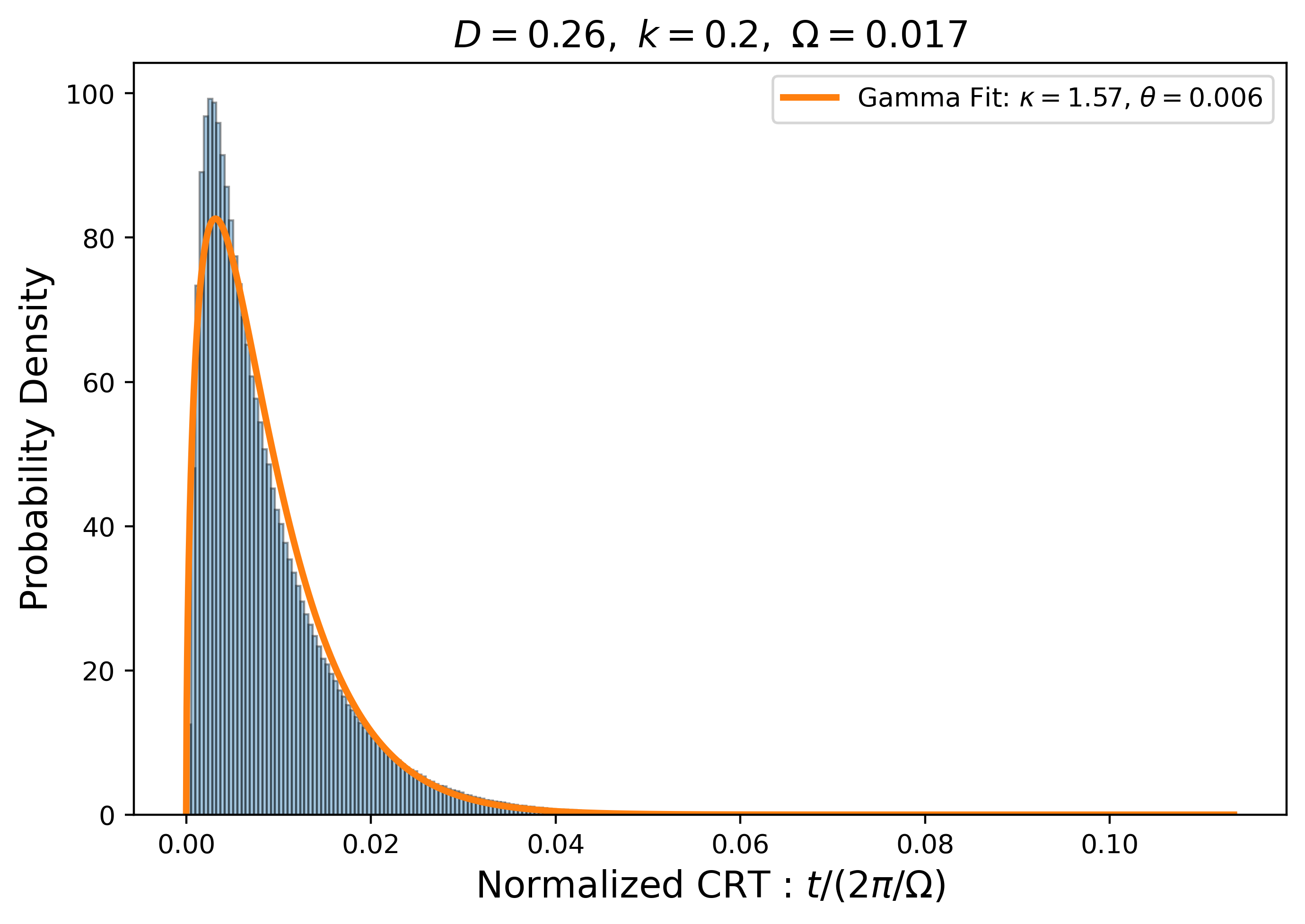}
        \caption{$D=0.26$}
        \label{fig:td1img3}
    \end{subfigure}
    \caption{Conditional residence time distribution for $k=0.2$ and $\Omega=0.017$}
    \label{fig:t1}
\end{figure}

\begin{enumerate}
    \item $\tau_{K}<T_{\Omega}/2$: In this case, the leader does not wait for the barrier to get minimized and escapes before the optimum condition is achieved. The follower then waits for the barrier to get minimized. If the $\tau_K$ for the follower is less than the time remaining for the barrier to get minimized the follower also crosses the barrier. On the other hand if the barrier gets minimized first, the condition becomes increasingly suitable for the follower to cross the barrier. In both cases, the CRT for the follower obeys the gamma distribution. This kind of statistics is mostly observed for intermediate and strong coupling. The CRTs depend not only on the noise strength but also on the coupling between the particles. The dependence on the coupling strength is observed to be much more significant than on the noise strength as shown in fig.\ref{fig:t1}-\ref{fig:t2}. For weak coupling, this statistics is observed for very small $\Omega$ or large $T_{\Omega}$. In such cases for the weak coupling, the gamma distribution is reduced to exponential distribution as the leader and follower behave almost like independent particles and the dependence on the noise strength dominates. This can be seen in fig. \ref{fig:t3}.
   
\begin{figure}[H]
    \centering
    \begin{subfigure}[b]{0.3\textwidth}
        \centering
        \includegraphics[width=\textwidth]{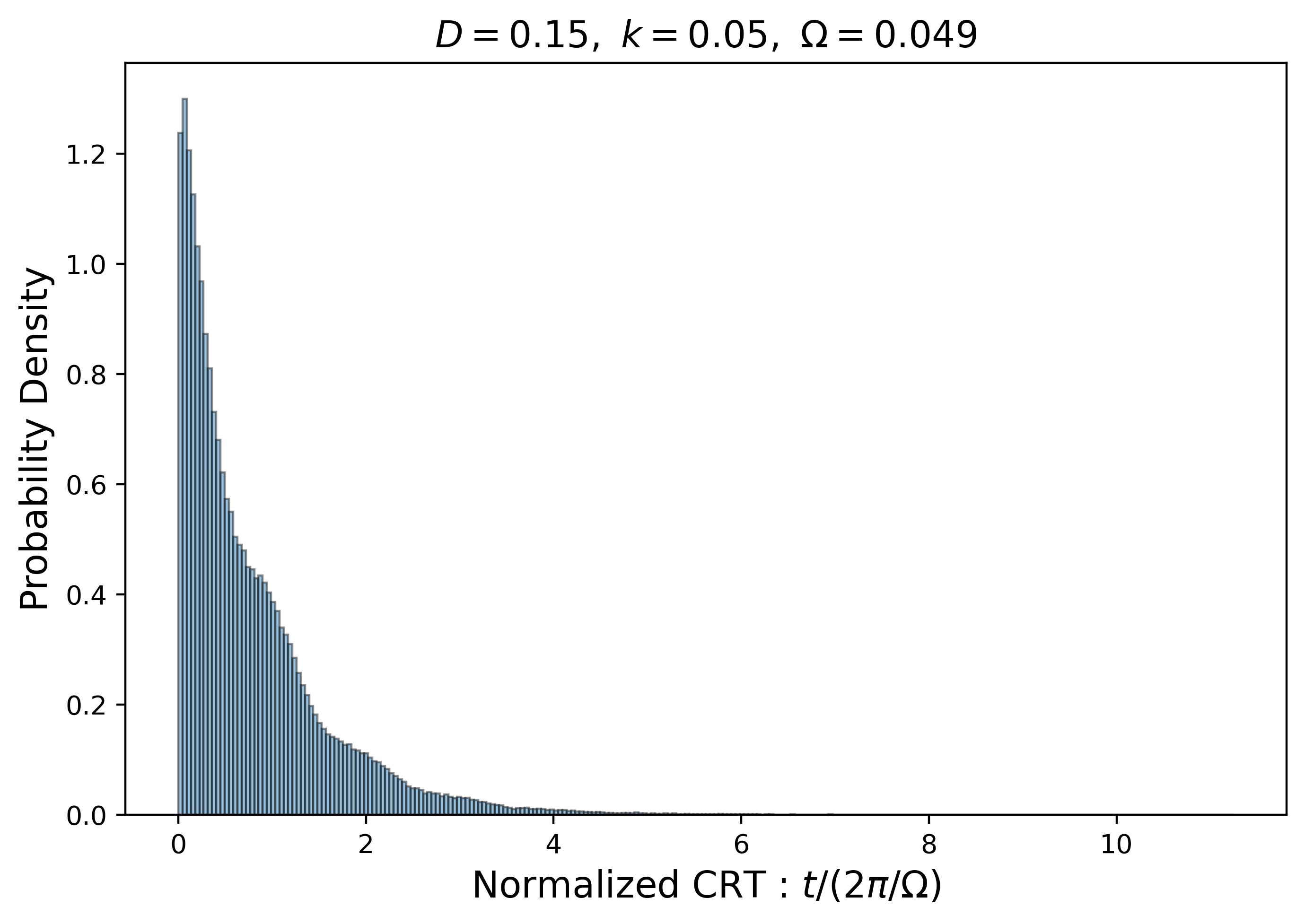}
        \caption{$D=0.15$}
        \label{fig:td3img1}
    \end{subfigure}
    \hfill
    \begin{subfigure}[b]{0.3\textwidth}
        \centering
        \includegraphics[width=\textwidth]{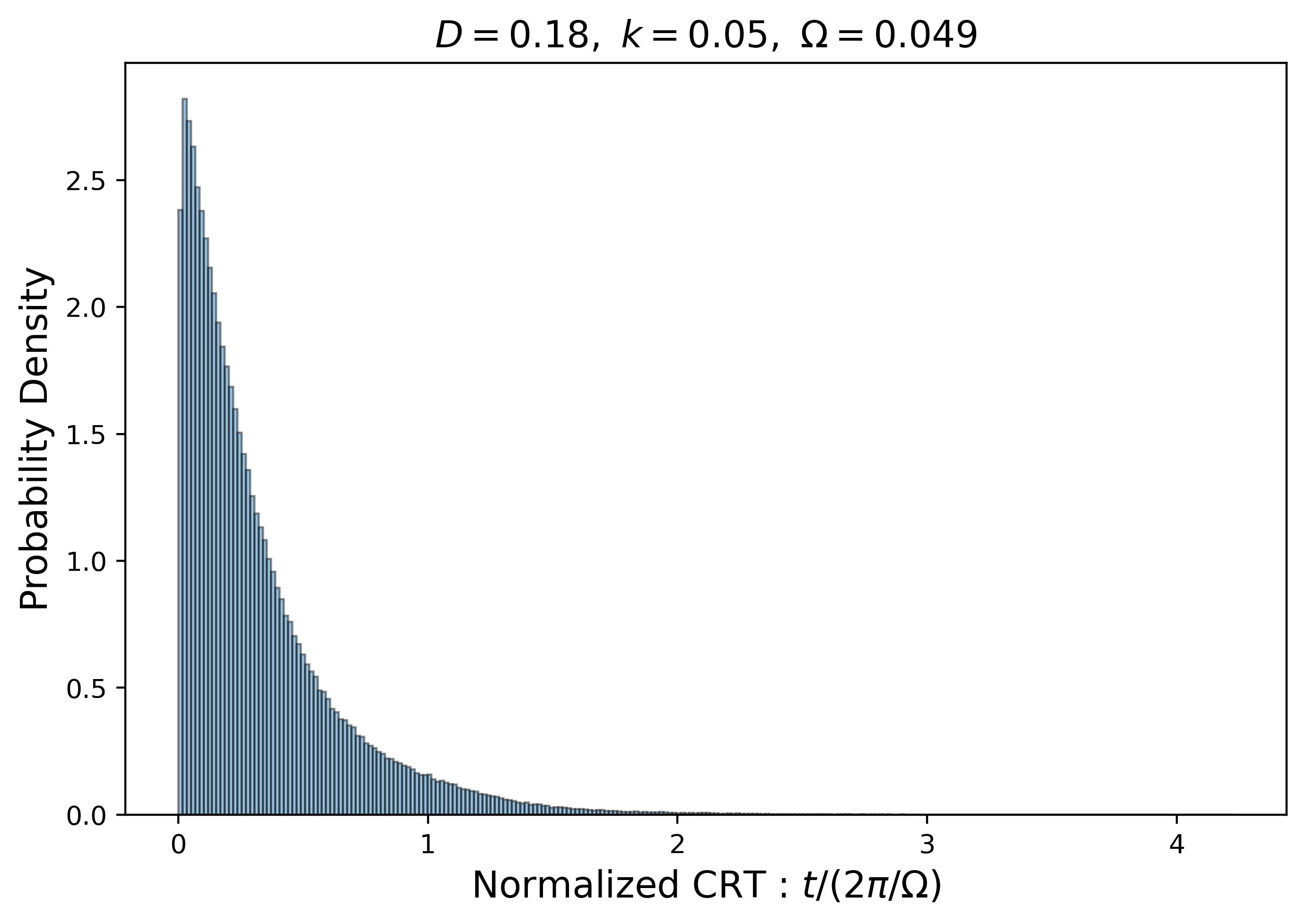}
        \caption{$D=0.18$}
        \label{fig:td3img2}
    \end{subfigure}
    \hfill
    \begin{subfigure}[b]{0.3\textwidth}
        \centering
        \includegraphics[width=\textwidth]{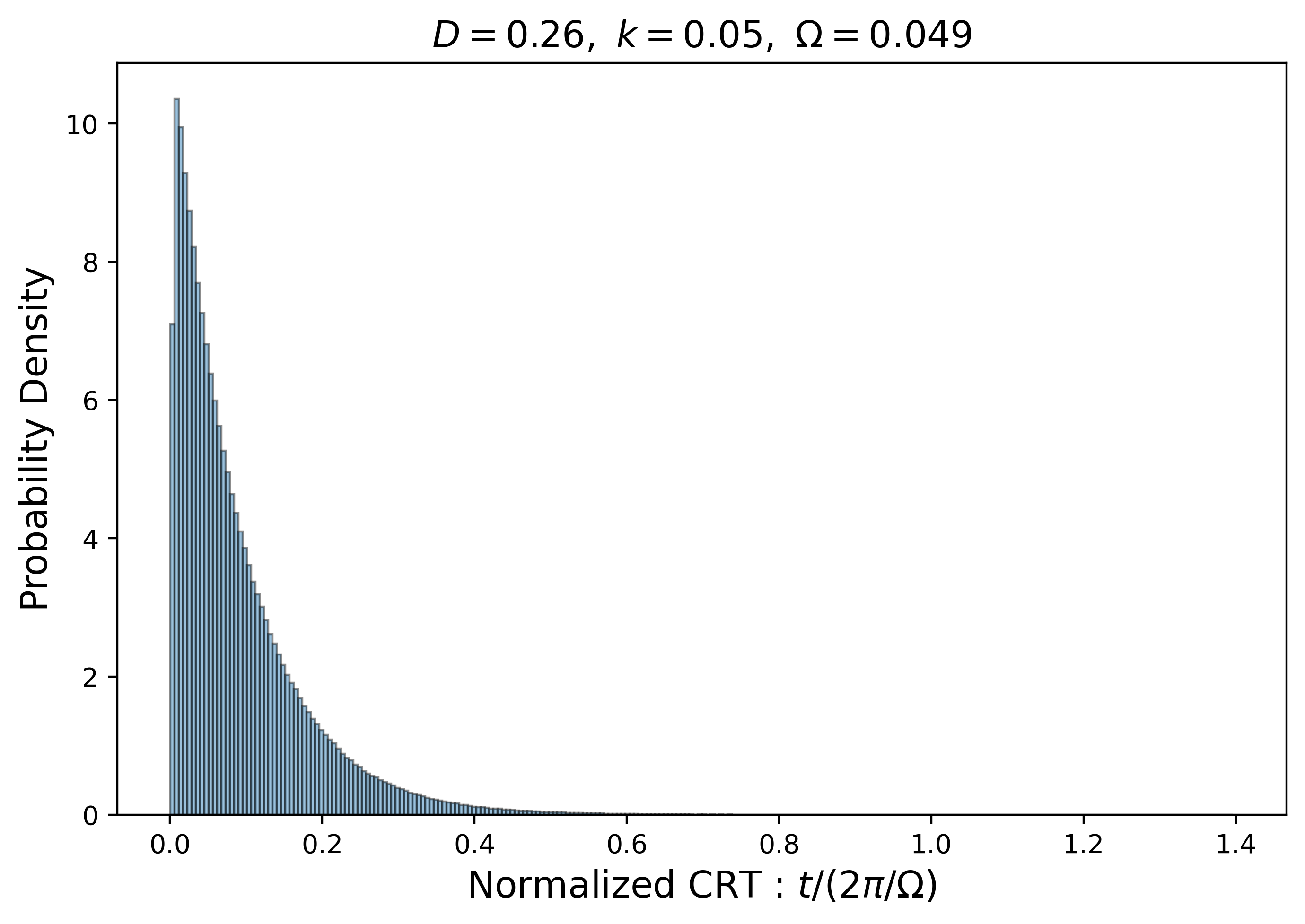}
        \caption{$D=0.26$}
        \label{fig:td3img3}
    \end{subfigure}
    
    \caption{Conditional residence time distribution for $k=0.05$ and $\Omega=0.049$.}
    \label{fig:t3}
\end{figure}

    \item $\tau_{K}\gtrsim T_{\Omega}/2$: When the two timescales are approximately equal, the leader initiates the transition when the barrier height is minimized. For intermediate and strong couplings, the CRTs are short and they take advantage of this favorable potential landscape and cross the barrier. However, for weak coupling there is still a non-negligible chance that the follower misses this opportunity. If so, then it has to wait a full time period $T_{\Omega}$ for the optimum condition to reappear. This is evident from the initial exponential structure in fig. \ref{fig:t3} followed by successive subtle bumps coinciding with multiples of $T_{\Omega}$. The heights of these peaks decreases exponentially. This behavior is similar to that observed in RTDs of a single particle. The only difference here is that since its a one way transition that is going to take place the peaks do not occur at odd multiples of $T_{\Omega}/2$. The number of successive peaks increases as $T_{\Omega}$ becomes smaller. The probability that the follower will miss the optimum condition increases as the barrier fluctuates more rapidly. The follower may have to wait over several time periods to make the transition. The similarity between the nature of CRTDs for weak coupling and RTDs for a single particle confirm the almost independently behaving nature of the two particles.

\end{enumerate}
Next, we have calculated the probabilities of escape of the follower. We have divided the whole process into three regions or windows:
\begin{enumerate}
        \item Zeroth window: This region includes the exponential structure of the distribution for half a time period just after the leader has crossed the threshold, i.e., $0<(t/T_{\Omega})<0.5$. 
        \item First window: This region includes the first peak if present in the distribution. This region includes the times when the follower has waiting almost 1 time period after the leader has escaped, i.e., $<0.5(t/T_{\Omega})<1.5$.
        \item Later window: This region includes the successive peaks after the first one, i.e., $(t/T_{\Omega})>1.5$.
    \end{enumerate}

    \begin{figure}[H]
    \centering
    \includegraphics[width=0.65\textwidth]{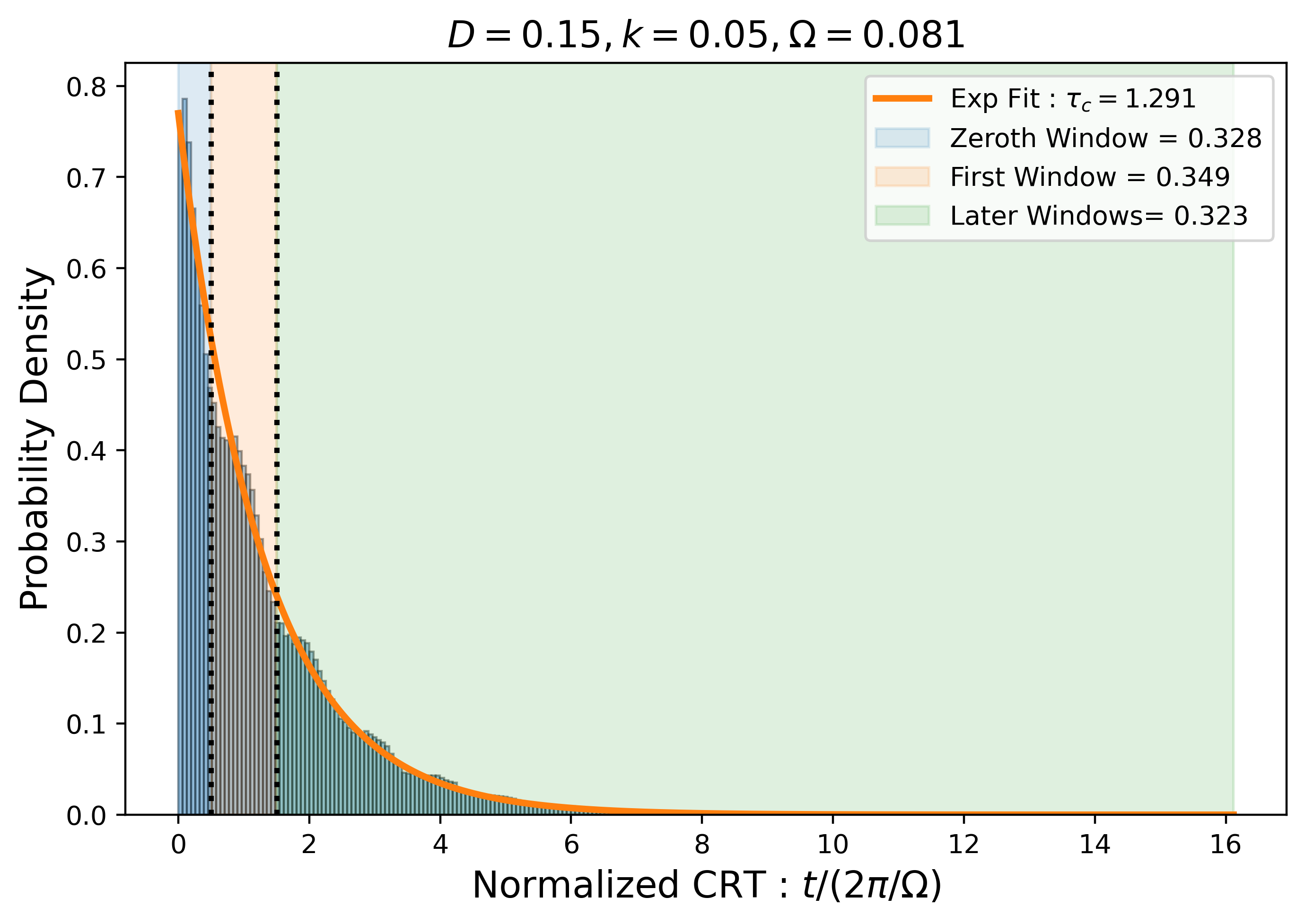}
    \caption{CRT distribution showing different windows of transition}        \label{fig:crtw}
\end{figure}
    Fig. \ref{fig:crtw} shows the different regions of transition. We have calculated the probability that the follower makes the transition in each region by varying $\Omega,D\,\text{and}\,k$. For intermediate and strong coupling the follower has a probability ranging from $0.9-0.99$ that it will utilize the zeroth window to make the transition, the remaining being the probability that the transition takes place in the first window. However, for weak coupling the probabilities change significantly. Table \ref{tab:window_probabilities} shows these probabilities. We see that since $\tau_K$ for the follower depends mostly on $D$, the probability of transition in the later windows is much higher for smaller values of noise, eventually becoming negligible as $D$ increases.

    \begin{table*}[t]
\centering
\caption{
Window-wise transition probabilities for different driving frequencies
$\Omega$ and noise strengths $D$ for $k=0.05$.
}
\label{tab:window_probabilities}

\renewcommand{\arraystretch}{1.15}

\begin{tabular}{c ccc ccc ccc}
\hline\hline

$\Omega$
& \multicolumn{3}{c}{$D=0.15$}
& \multicolumn{3}{c}{$D=0.18$}
& \multicolumn{3}{c}{$D=0.26$} \\

\cline{2-4}
\cline{5-7}
\cline{8-10}


& \shortstack{Zeroth \\ $(P_0)$}
& \shortstack{First \\ $(P_1)$}
& \shortstack{Later \\ $(P_L)$}

& \shortstack{Zeroth \\ $(P_0)$}
& \shortstack{First \\ $(P_1)$}
& \shortstack{Later \\ $(P_L)$}

& \shortstack{Zeroth \\ $(P_0)$}
& \shortstack{First \\ $(P_1)$}
& \shortstack{Later \\ $(P_L)$} \\

\hline

0.001 & 1.000 & 0.000 & 0.000 & 1.000 & 0.000 & 0.000 & 1.000 & 0.000 & 0.000 \\
0.009 & 0.970 & 0.030 & 0.000 & 1.000 & 0.000 & 0.000 & 1.000 & 0.000 & 0.000 \\
0.017 & 0.849 & 0.147 & 0.004 & 0.987 & 0.013 & 0.000 & 1.000 & 0.000 & 0.000 \\
0.025 & 0.722 & 0.254 & 0.023 & 0.949 & 0.051 & 0.000 & 1.000 & 0.000 & 0.000 \\
0.033 & 0.628 & 0.314 & 0.058 & 0.896 & 0.102 & 0.001 & 1.000 & 0.000 & 0.000 \\
0.041 & 0.547 & 0.352 & 0.101 & 0.844 & 0.151 & 0.005 & 0.999 & 0.001 & 0.000 \\
0.049 & 0.476 & 0.373 & 0.151 & 0.793 & 0.196 & 0.010 & 0.997 & 0.003 & 0.000 \\
0.057 & 0.431 & 0.369 & 0.200 & 0.732 & 0.248 & 0.020 & 0.993 & 0.007 & 0.000 \\
0.065 & 0.371 & 0.384 & 0.245 & 0.683 & 0.285 & 0.032 & 0.987 & 0.013 & 0.000 \\
0.073 & 0.337 & 0.378 & 0.284 & 0.644 & 0.308 & 0.048 & 0.979 & 0.021 & 0.000 \\
0.081 & 0.328 & 0.349 & 0.323 & 0.601 & 0.335 & 0.064 & 0.970 & 0.030 & 0.000 \\
0.089 & 0.287 & 0.364 & 0.349 & 0.567 & 0.349 & 0.084 & 0.960 & 0.040 & 0.000 \\
0.097 & 0.264 & 0.352 & 0.384 & 0.525 & 0.375 & 0.100 & 0.948 & 0.052 & 0.000 \\
0.105 & 0.251 & 0.328 & 0.421 & 0.505 & 0.370 & 0.125 & 0.936 & 0.064 & 0.000 \\
0.113 & 0.248 & 0.305 & 0.447 & 0.474 & 0.385 & 0.140 & 0.918 & 0.081 & 0.001 \\
0.121 & 0.226 & 0.311 & 0.463 & 0.465 & 0.371 & 0.164 & 0.908 & 0.091 & 0.001 \\
0.129 & 0.196 & 0.308 & 0.496 & 0.437 & 0.383 & 0.180 & 0.889 & 0.110 & 0.001 \\
0.137 & 0.200 & 0.288 & 0.512 & 0.412 & 0.382 & 0.206 & 0.873 & 0.125 & 0.002 \\
0.145 & 0.179 & 0.291 & 0.530 & 0.398 & 0.383 & 0.219 & 0.861 & 0.136 & 0.003 \\
0.153 & 0.188 & 0.275 & 0.537 & 0.383 & 0.381 & 0.236 & 0.845 & 0.151 & 0.004 \\
0.161 & 0.175 & 0.263 & 0.561 & 0.378 & 0.367 & 0.255 & 0.831 & 0.164 & 0.005 \\
0.169 & 0.149 & 0.266 & 0.585 & 0.362 & 0.363 & 0.274 & 0.818 & 0.175 & 0.006 \\

\hline\hline
\end{tabular}
\end{table*}

\subsubsection{Role of $\Omega$}
\begin{figure}[H]
    \centering
    \includegraphics[width=0.4\textwidth]{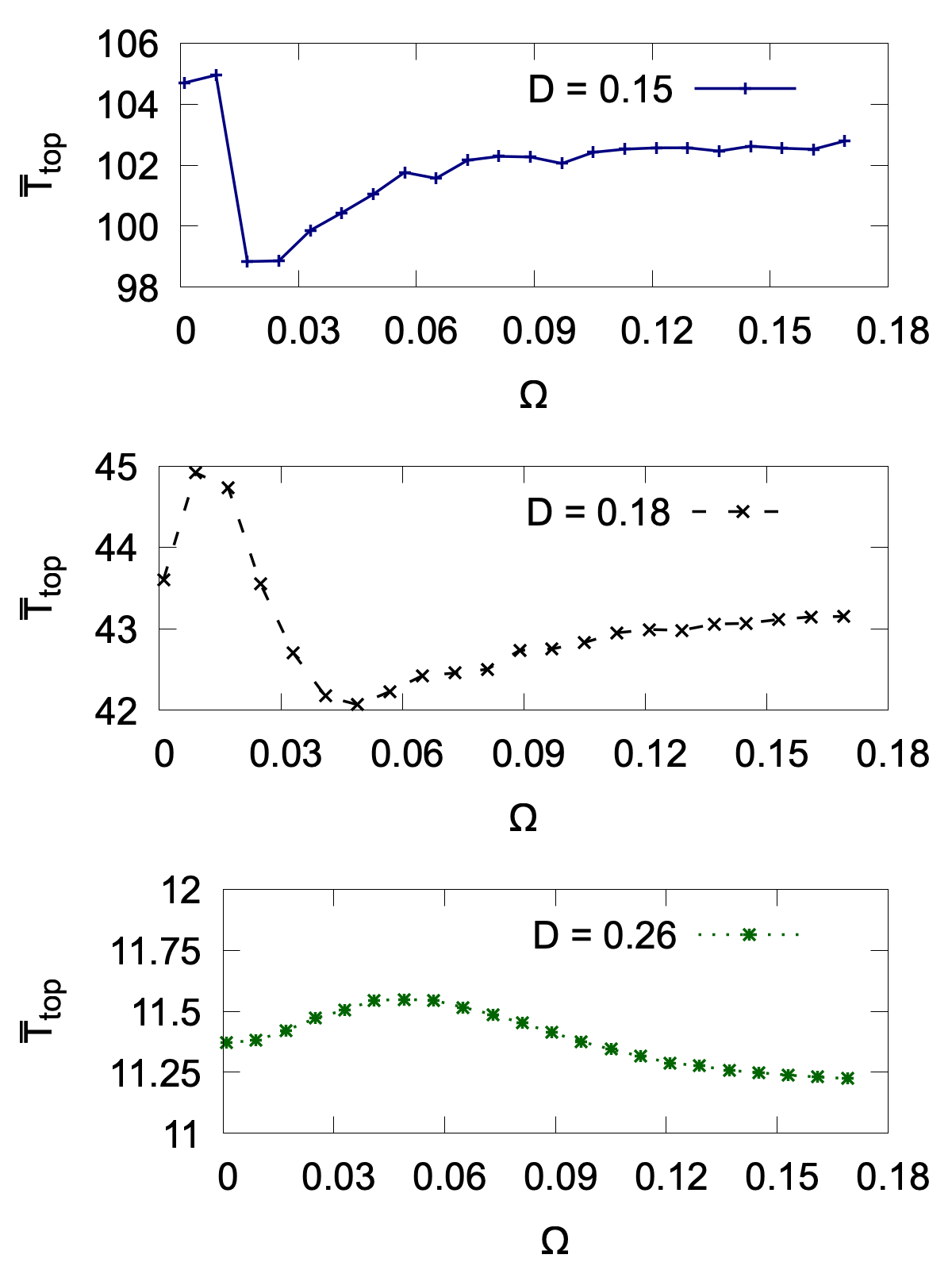}
    \caption{CRT vs $\Omega$ for $k=0.05$}        \label{fig:ttimg1}
\end{figure}

\begin{figure}[H]
    \centering
    \includegraphics[width=0.4\textwidth]{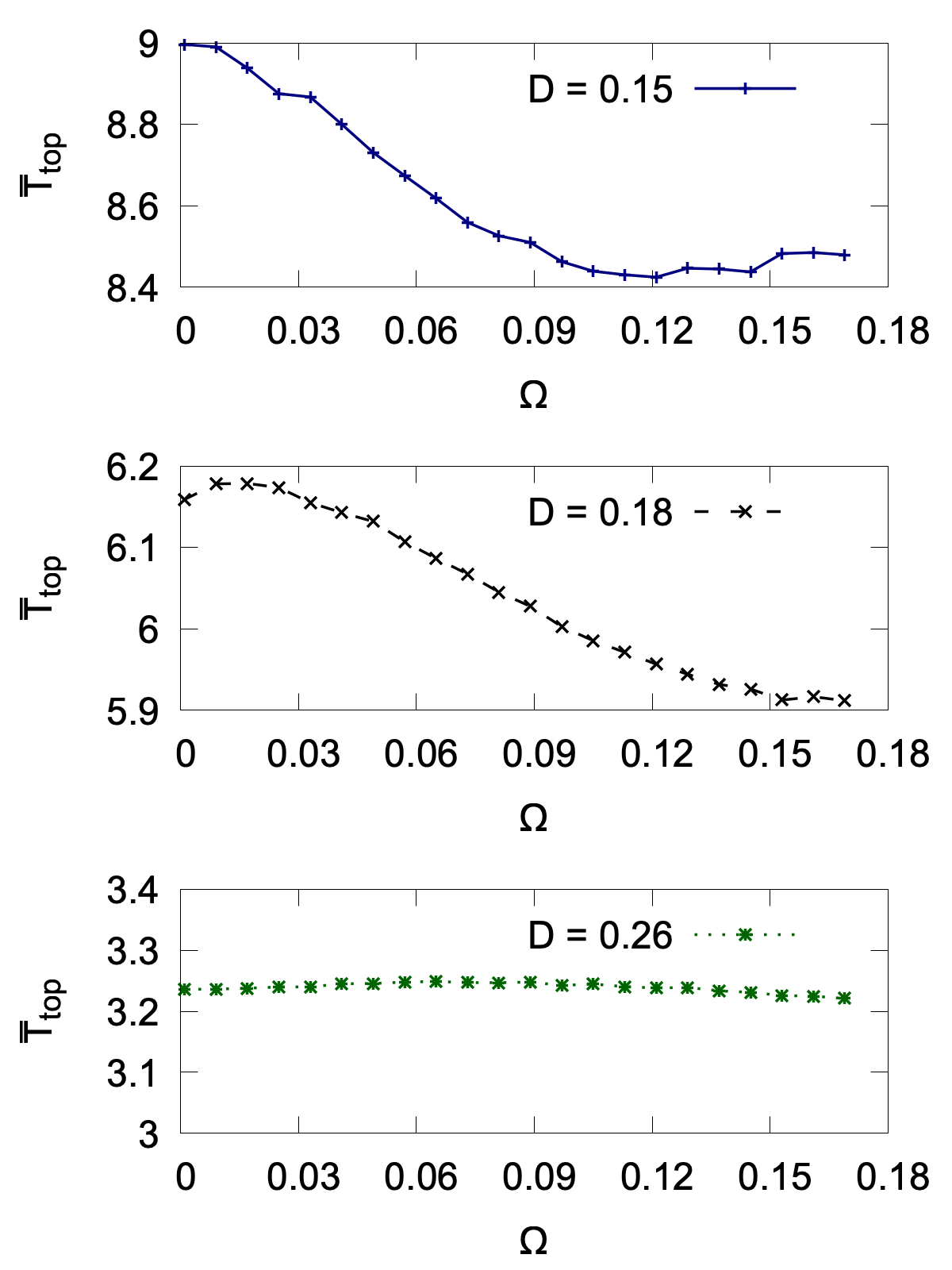}
    \caption{CRT vs $\Omega$ for $k=0.2$}        \label{fig:ttimg2}
\end{figure}

\begin{figure}[H]
    \centering
    \includegraphics[width=0.4\textwidth]{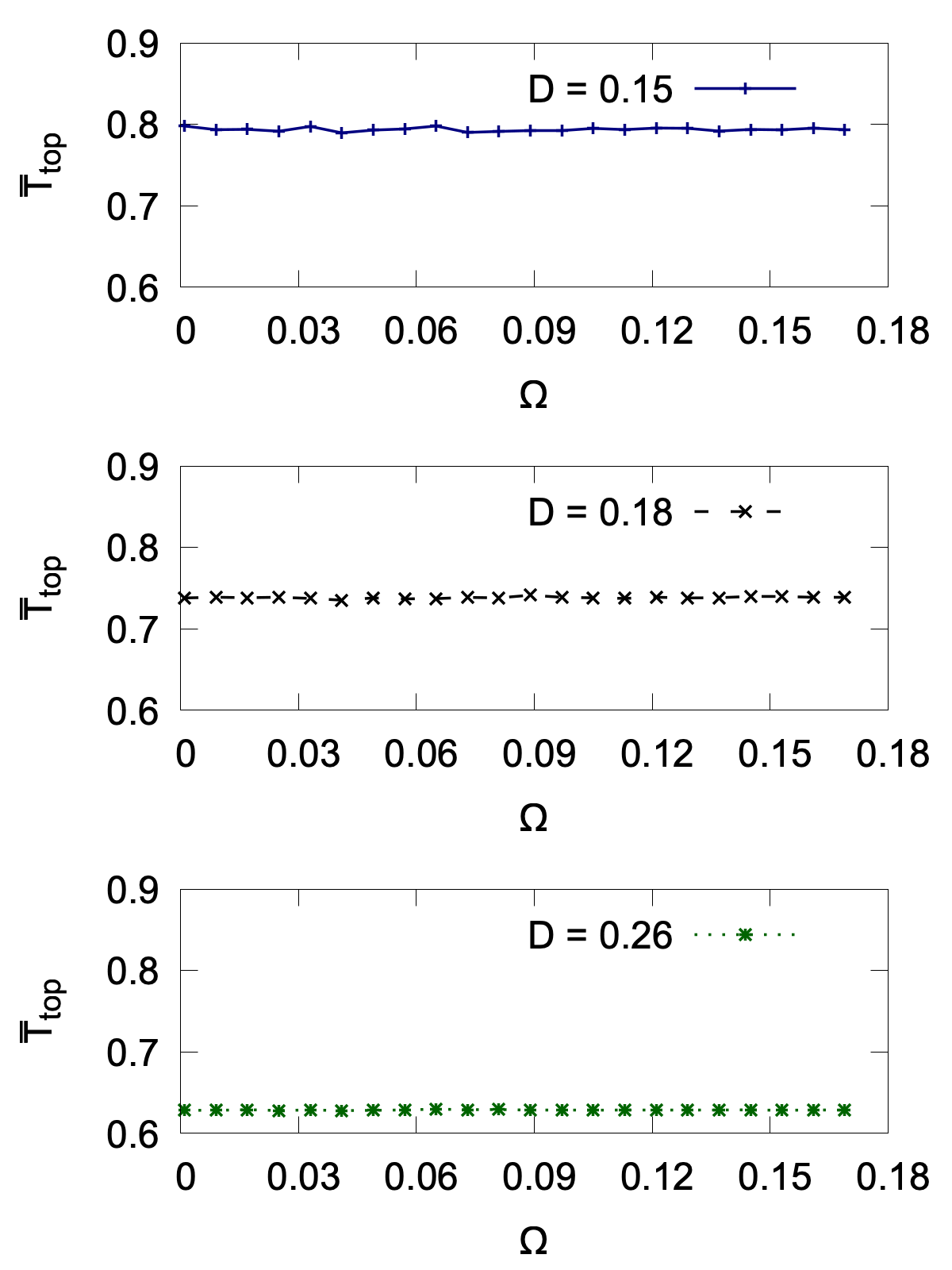}
    \caption{CRT vs $\Omega$ for $k=1.0$}        \label{fig:ttimg3}
\end{figure}

An important aspect to be noted from the studies of CRTs is its dependence on the frequency of the periodic modulation. The time-scales associated with the catch-up process are much smaller than the ones linked with the complete transition of the dimer across the barrier. The CRT distributions show that the most pronounced enhancements to this catch-up process are brought about by drives having slow to moderate frequencies, that are still greater than the frequency used for obtaining SR in \cite{62t9-g2n6}. This is reflected in the mean CRT plots in figures \ref{fig:ttimg1}-\ref{fig:ttimg3}. 

The dependence of the mean conditional residence time on the driving frequency provides further insight into the completion dynamics discussed in the previous sections. While the CRT distributions reveal how the probability of transition is distributed across different synchronization windows, the mean CRT summarizes the overall delay associated with the catch-up process. The variation of the mean CRT with $\Omega$, shown in Figs. \ref{fig:ttimg1}-\ref{fig:ttimg3}, therefore reflects how the redistribution of transition probabilities across the zeroth, first, and later windows modifies the average completion time of the transition.
For weak coupling (k=0.05), the mean CRT decreases significantly with increasing $\Omega$ for small and intermediate noise strengths. This behavior is consistent with the CRT distributions and the window-wise probabilities discussed earlier. 
At small frequencies, i.e. large time periods, the escape of the follower is mostly due to the noise induced escape process as the barrier fails to get minimized before the follower escapes. As we increase the frequency, the time period gets closer to the value of the escape time of the follower and hence we see a reduction in the mean CRT values. From the CRT distributions and \ref{tab:window_probabilities}, we can see the probability distribution shifts from predominantly zeroth-window events to a mixture of zeroth- and first-window transitions. This is the region where we get the minimum in the mean CRT 

\noindent plots. As we increase the frequency further, we observe that the follower increasingly misses the chance to make the transition in the initial windows and hence waits for more number of cycles. However, the mean CRT does not change appreciably. This is because although the transition occurs over multiple cycles, the actual time period has decreased. The balance between the two causes the mean CRT to not show significant changes.

For intermediate coupling (k=0.2), a similar but weaker trend is observed. In this regime, the CRT distributions already exhibit a stronger dominance of the zeroth window and relatively smaller later-window contributions. Although the probabilities of the first and later windows still increase with $\Omega$, their growth remains comparatively limited because the stronger coupling enables the follower to complete the transition more efficiently once initiated. As a result, the redistribution of probability is weaker than in the soft-dimer case, producing a more gradual decrease in the mean CRT. The shallow minima observed in the

\noindent corresponding plots are therefore associated with a comparatively smaller redistribution of transition probability toward the first synchronization window. 

In contrast, for strong coupling $(k=1.0)$, the mean CRT remains nearly independent of $\Omega$. This behavior directly follows from the CRT distributions discussed previously, where

\noindent the transition probability remains overwhelmingly concentrated within the zeroth window throughout the studied frequency range. Since delayed transitions over one or multiple forcing periods are strongly suppressed in the rigid-dimer regime, increasing $\Omega$ produces only a very small increase in the first-window probability and an almost negligible contribution from the later windows. Consequently, the overall temporal structure of the completion process changes very little with frequency, resulting in the nearly flat mean CRT curves observed in Fig. \ref{fig:ttimg3}. 

 The CRT versus $\Omega$ plots must therefore be interpreted together with the corresponding CRT distributions and window-wise probabilities. The minima in the mean CRT curves are not simply associated with faster transitions, but rather with the frequency range where the probability of transitions in the zeroth window shifts to a combination of transitions in zeroth and the first window. This demonstrates that the external drive reorganizes the completion dynamics by shifting transitions from long delayed events distributed over several forcing cycles toward completion within the first synchronization window.
 
\section{Discussion and Conclusion}\label{sec5}
In this study, the completion dynamics across the barrier of an overdamped dimer has been explored. Unlike the conventional quantifiers like the residence times, escape rates or the mean first passage times, our study is focused on the delay between the initiation and the completion of a successful transition. We have used the Conditional Residence Time (CRT), defined as the time interval between the threshold crossing of the leader and the subsequent crossing of the follower, as a quantifier.

The CRT distributions show that the completion dynamics is dependent on the competition between the escape time of the follower and the time period of the external drive. This leads to qualitatively different transition pathways. When the follower is able to respond within the very first favorable phase of the drive the CRT distributions follow the Gamma distribution. When the follower misses this chance it has to wait for some time, leading to its transitions being distributed over several cycles of the drive. This is reflected as a multipeak structure embedded on the gamma like distribution.

The coupling strength plays a strong role in deciding which pathway will be taken. For strong and intermediate coupling, the probability that the follower escapes after a short delay, i.e. without skipping even a single cycle of the drive, provided the noise strength is sufficient. The leads to either no peaks or very faint peaks in the distributions. The distributions show very weak sensitivity to the frequency of the external drive. For weak coupling, the two monomers behave almost independently and hence the delayed completion occurs with a significant probability.

In order to characterize this redistribution of probability over different cycles of the drive, we have presented a window-wise description of the completion process. We study three distinct windows, i.e. zeroth, first and the later windows corresponding to transitions occurring immediately after the initiation, after one cycle of the drive and in later cycles respectively. As the frequency is increased the probability of transition of the follower shifts to the first and the later windows. This redistribution of the probability also explains the non-monotonic structure of the mean CRT presented in the work. The mean CRT shows a non-monotonic dependence of the frequency of the external drive. This is related to where the probability shifts from the transition occurring in just the zeroth window to the transitions occurring in both zeroth and the first window. This dependence weakens as the coupling strength is increased as the transitions then are mostly controlled by the strong restoring forces and the dynamics shifts from being a drive controlled one to the coupling controlled one. 

Our study shows that the transition initiation and the completion are two distinct processes that may occur at different timescales. The CRT provides a measure that directly probes the cooperative dynamics occurring during the catch-up process. More generally, the CRT framework may be useful for investigating sequential escape processes in polymers, coupled oscillators, molecular systems, and other interacting stochastic systems where one unit of the system initiates the transition but the completion depends on the response of the remaining units of the chain.

\section*{Acknowledgments}
\noindent W.L.R. acknowledges Prof. M. C. Mahato for the insightful discussion, and Anusandhan National Research Foundation (ANRF) (Project No. EEQ/2021/000434) for providing the financial assistance.

\bibliography{manu}

\end{document}